\documentclass[aps,prl,reprint,superscriptaddress,showpacs]{revtex4-1}

\usepackage{graphicx}

\newcommand{\CCO}{CuCrO$_{2}$}

\begin{document}

\title{Development of spin correlations in the geometrically
frustrated triangular-lattice Heisenberg antiferromagnet CuCrO$_{2}$}

\author{R. Kajimoto}
\email[E-mail: ]{ryoichi.kajimoto@j-parc.jp}
\affiliation{Research Center for Neutron Science and Technology, CROSS,
Tokai, Ibaraki 319-1106, Japan}
\affiliation{J-PARC Center, Japan Atomic Energy Agency, Tokai, Ibaraki
319-1195, Japan}

\author{K. Tomiyasu}
\affiliation{Department of Physics, Tohoku University, Sendai 980-8578,
Japan}

\author{K. Nakajima}
\affiliation{J-PARC Center, Japan Atomic Energy Agency, Tokai, Ibaraki
319-1195, Japan}

\author{S. Ohira-Kawamura}
\affiliation{J-PARC Center, Japan Atomic Energy Agency, Tokai, Ibaraki
319-1195, Japan}

\author{Y. Inamura}
\affiliation{J-PARC Center, Japan Atomic Energy Agency, Tokai, Ibaraki
319-1195, Japan}

\author{T. Okuda}
\affiliation{Department of Nano-Structures and Advanced Materials,
Kagoshima University, Kagoshima 890-0065, Japan}

\date{\today}

\begin{abstract}
 Magnetic excitations in the triangular-lattice Heisenberg
 antiferromagnet (TLHA) {\CCO} were studied using single-crystal
 inelastic neutron scattering. A diffusive quasielastic component that
 persisted without developing a correlation length over a wide
 temperature range both below and above the ordering temperature was
 observed. Furthermore, characteristic momentum dependence was observed
 that was reproduced using minimum spin clusters. The robust spin
 clusters contrast with conventional magnetic ordering and may be
 universal in TLHAs.
\end{abstract}

% We investigated temperature dependence of the magnetic excitations in
% the triangular-lattice Heisenberg antiferromagnet (TLHA) {\CCO} by
% single-crystal inelastic neutron scattering. We found that the
% excitation spectrum has a diffusive quasielastic component at finite
% temperatures, and it shows characteristic momentum dependence
% consistent with scattering by spin clusters of minimum triangular
% units of the 120$^{\circ}$ structure. The diffuse component crossovers
% to spin wave as the temperature decreases without developing the
% correlation length. This fact suggests that the ordering of spins in
% TLHA cannot be described only by the divergence of the correlation
% length like a conventional magnet, but it is accompanied by the
% decrease of number of the spin clusters.

% 75.25.-j Spin arrangements in magnetically ordered materials
% (including neutron and spin-polarized electron studies,
% synchrotron-source x-ray scattering, etc.)
% 75.40.Gb Dynamic properties (dynamic susceptibility, spin waves, spin
% diffusion, dynamics scaling, etc.)
% 75.47.Lx Magnetic oxides

\maketitle

%\section{Introduction}

A two-dimensional (2D) triangular-lattice Heisenberg antiferromagnet
(TLHA) is a typical and one of the simplest examples of geometrically
frustrated antiferromagnets, in which a novel spin state originating
from competing magnetic interactions and low dimensionality is
expected. Although the study of this type of system originated with the
resonating valence bonds theoretically predicted by Anderson more than
four decades ago \cite{anderson73}, the realization of this novel state
was not confirmed experimentally until recently, and this system still
remains of great interest in condensed matter physics. The spin liquid
state was first confirmed in the organic $S=1/2$ systems,
$\kappa$-(BEDT-TTF)$_{2}$Cu$_{2}$(CN)$_{3}$ \cite{shimizu03} and
EtMe$_{3}$Sb[Pd(dmit)$_{2}$]$_{2}$ \cite{itou08,itou10}, for which the
quantum fluctuations of $S=1/2$ spins prevent antiferromagnetic spin
ordering, even at zero temperature ($T$). Furthermore, other novel spin
states can be realized in larger $S$ systems with short-range spin
correlations. For example, thermodynamic and powder neutron scattering
studies of NiGa$_{2}$S$_{4}$ revealed that this material shows a low-$T$
disordered state of $S=1$ spins with short-range correlations, and this
behavior was interpreted as the formation of a spin liquid state
\cite{nakatsuji05}. Furthermore, an $S=3/2$ system, NaCrO$_{2}$, for
which the classical nature of the spins should dominate, shows an
unconventional fluctuating crossover regime at \emph{finite} $T$ below
the spin transition temperature $T_c \sim 40$~K \cite{olariu06}. A
powder neutron scattering study showed that the spin correlations induce
diffuse quasielastic scattering in this $T$ region \cite{hsieh08}. This
spin fluctuation is speculated to be an evidence of excitations of $Z_2$
vortices \cite{kawamura84,okubo10,kawamura11}, although its origin has
not yet been identified.

%whose appearance in a TLHA is predicted in a regime where 2D spin
%correlations dominate

The delafossite oxide {\CCO} is a 2D TLHA. This compound is similar to
the ordered rock-salt compound NaCrO$_{2}$ in that the $S = 3/2$ spins
of the Cr$^{3+}$ (3d$^{3}$) ions form a 2D triangular lattice, and the
Cr layers stack in a rhombohedral manner ($ABCABC\cdots$)
\cite{seki08}. In this compound, 2D spin correlations begin to develop
around the Curie-Weiss temperature $T_\mathrm{CW} = 160$--200~K
\cite{poienar09,frontzek11,doumerc86,okuda05,okuda09}. Because of finite
inter-layer couplings, three-dimensional (3D) ordering of a nearly
120$^{\circ}$ structure occurs below $T_N \sim 24$~K
\cite{kadowaki90,poienar09,soda09}. However, the magnetic specific heat
($C_\mathrm{mag}$) exhibits a broad shoulder structure in addition to
the sharp peak for the 3D spin ordering around $T_N$ \cite{okuda08},
suggesting the existence of additional spin fluctuations. Although the
sharp peak is suppressed by the substitution of the Cu$^{+}$ (Cr$^{3+}$)
ion with a Ag$^{+}$ (Al$^{3+}$) ion reflecting the suppression of the 3D
ordering, the broad shoulder structure remains, which together with the
$T^2$ dependence of $C_\mathrm{mag}$ at low $T$ suggests unconventional
2D low-energy spin fluctuations \cite{okuda09,okuda11}. Concomitantly
with these anomalies in $C_\mathrm{mag}$, diffuse quasielastic
scattering similar to that observed in NaCrO$_{2}$ is commonly observed
via neutron scattering \cite{kajimoto10,kajimoto13}, confirming the
existence of dynamical 2D short-range spin correlations. However,
because the neutron scattering studies of these Cr compounds have been
limited to powder samples, detailed information, particularly regarding
the spatial (momentum) distribution of the spin fluctuations, has not
yet been obtained.

Accordingly, in this Letter, we report an inelastic neutron scattering
(INS) study of a single crystal of {\CCO} for detailed elucidation of
the diffuse component in the magnetic excitation spectrum observed for
powder samples.  Close investigation revealed a unique $Q$ and $T$
dependence of the diffuse component. The origin of this correlation and
its relationship to the development of spin correlations in TLHA are
also discussed.

\begin{figure*}[t]
 \includegraphics[scale=0.5]{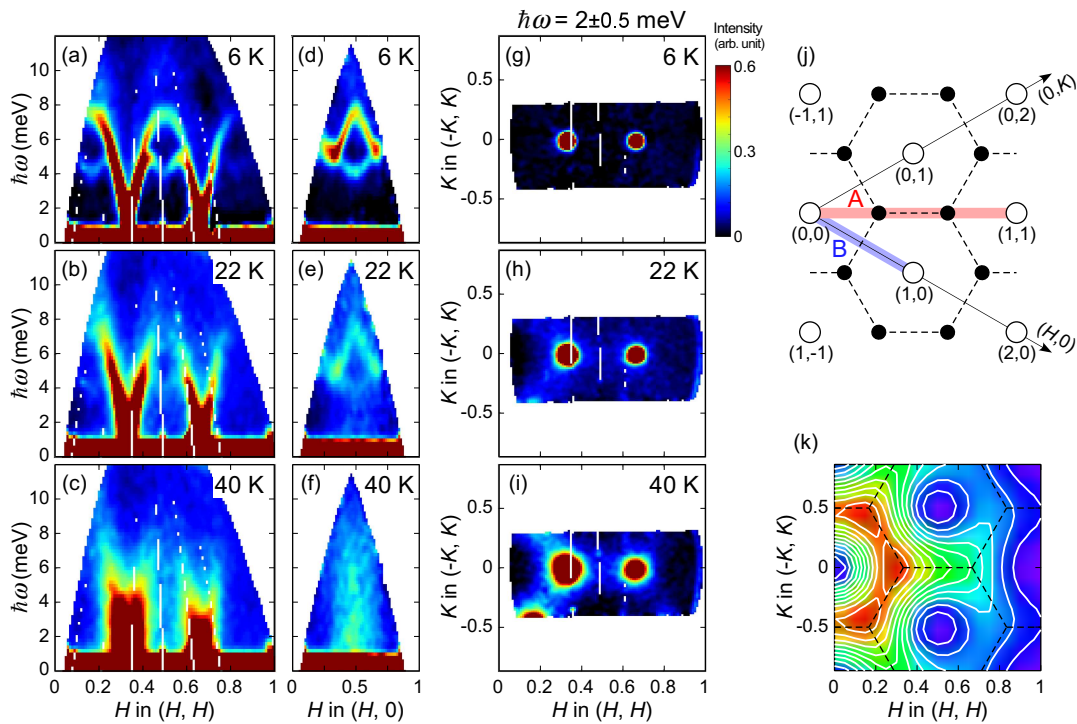}
 \caption{(Color online) (a)--(i) Excitation spectra of {\CCO} at 6, 22,
 and 40~K. (a)--(c) The $Q$-$\hbar\omega$ maps cut along $(H,H)$ [line A
 in (j)] with widths of $K=0 \pm 0.04$ in $(-K,K)$. (d)--(f) The
 $Q$-$\hbar\omega$ maps cut along $(H,0)$ [line B in (j)] with widths of
 $K=0 \pm 0.02$ in $(-K,2K)$. (g)--(i) Maps on the $(H,H)$-$(-K,K)$
 planes cut at $\hbar\omega = 2$~meV with widths of $\pm$0.5~meV. (j) 2D
 reciprocal lattice of {\CCO}. Open circles, closed circles, and broken
 lines show $\Gamma$ points, K points, and Brillouin zone boundaries,
 respectively. A and B show the directions of the cuts of data (see
 text). (k) Neutron scattering intensity map calculated based on the
 trimer model. Broken lines show Brillouin zone boundaries.}
 \label{spectra}
\end{figure*}

%\section{Experiments}

A 140~mg single crystal of {\CCO} was synthesized by a flux method. The
crystal structure was rhombohedral (space group $R\bar{3}m$) with
lattice constants of $a \sim 3.0$~{\AA} and $c \sim 17$~{\AA} in a
hexagonal setting \cite{okuda09}. The INS measurement was performed
using the Cold-Neutron Disk-Chopper Spectrometer AMATERAS at J-PARC
\cite{nakajima11}. The incident energy of the neutrons and the energy
resolution at elastic scattering were 15~meV and 1~meV,
respectively. All the data were analyzed using the software suite
\textsc{utsusemi} \cite{inamura10} to obtain the dynamical structure
factor $S(\mathbf{Q},\omega)$ as described in \cite{kajimoto10}
($\mathbf{Q}$ and $\hbar\omega$ are the momentum and energy transfers,
respectively).  The crystal was initially aligned such that $\mathbf{c}$
was parallel to the incident neutron beam and the [110] direction was
horizontal, and then rotated about the vertical axis by
40$^{\circ}$. The $Q_z$ dependence of the data was ignored, because the
magnetic excitations were quasi 2D \cite{poienar10,frontzek11}, whereas
the $\mathbf{Q}$ dependence was represented by the 2D hexagonal
reciprocal lattice $(Q_x,Q_y) = (Ha^*,Ka^*)$ with $a^* =
4\pi/(\!\sqrt{3}a)$ [Fig.~\ref{spectra}(j)]. This configuration and
analysis enabled mapping of the 2D reciprocal space on the 2D detector
arrays of the instrument.  For the data at the lowest $T = 6$~K, the
data for an empty aluminum cell filled with helium gas were subtracted
to correct for the background due to scattering of the helium in the
sample cell. For the measurement of the magnetic Bragg peak
[Fig.~\ref{FitTdep}(a)], the monochromating chopper was halted in the
open position (diffraction mode). A preliminary result of the present
study was reported in \cite{kajimoto_sces}.

%\section{Results and Discussion}

First, the magnetic Bragg peak was measured to define the value of $T_N$
for the sample. Figure~\ref{FitTdep}(a) shows the $T$ dependence of the
integrated intensity of the magnetic Bragg peak $(1/3,1/3,0)$.  From
these data, the value for $T_N$ of the present sample was observed to be
24~K, which is the temperature at which the intensity shows a clear
increase with decreasing $T$.

Figures~\ref{spectra}(a)--(c) show the excitation spectra at 6, 22, and
40 K, which are cut along $(H,H)$ [line A in Fig.~\ref{spectra}(j)]. At
6~K, the spin-wave excitations dispersing from the magnetic Bragg peak
positions (K points), $\mathbf{Q} = (1/3,1/3)$ and $(2/3,2/3)$, were
observed over $\sim$8~meV [Fig.~\ref{spectra}(a)], and the dispersion
relation was consistent with previously reported results obtained at 2~K
\cite{frontzek11}. It should be noted that a continuum scattering with
less intensity existed above the spin wave dispersion over $\sim$10~meV,
which likely originated from two-magnon scattering \cite{mourigal13}. As
$T$ increased, the spectrum broadened [Figs.~\ref{spectra}(b) and
\ref{spectra}(c)]. Furthermore, a weak and diffusive signal filling the
$Q$--$\hbar\omega$ region between the dispersions from $(1/3,1/3)$ and
$(2/3,2/3)$ was detected. The appearance of the diffuse scattering can
be distinguished more clearly when the excitation spectrum is cut along
$(H,0)$ [line B in Fig.~\ref{spectra}(j)], as shown in
Figs.~\ref{spectra}(d)--(f). At 6~K, two branches of the spin wave
excitations were observed in the region $\hbar\omega = 4$--8~meV
[Fig.~\ref{spectra}(d)]. At 22~K, the diffuse scattering appeared at
approximately $H = 0.5$, connecting the spin wave branches and
$\hbar\omega = 0$~meV [Fig.~\ref{spectra}(e)]. It should also be noted
that, in this $T$ region slightly below $T_N$, the diffuse scattering
coexisted with the spin wave branches. At 40~K, the diffuse scattering
became more intense, whereas the spin wave branches were severely damped
[Figs.~\ref{spectra}(c) and \ref{spectra}(f)].

To examine the $\mathbf{Q}$ dependence of the diffuse scattering, the
data was cut at $\hbar\omega = 2$~meV. Figures~\ref{spectra}(g)--(i)
show the obtained spectra on the $(H,K)$ plane at 6, 22, and 40~K. At
6~K, two bright spots were observed at $(1/3,1/3)$ and $(2/3,2/3)$,
which originated from the spin waves [Fig.~\ref{spectra}(g)]. As $T$
increased to 22~K, the diffuse scattering appeared as ridges connecting
the spin-wave spots [Fig.~\ref{spectra}(h)]. Additionally, the intensity
of the diffuse scattering was pronounced as $T$ was further elevated to
40~K [Fig.~\ref{spectra}(i)].  Notably, the intensity of the diffuse
scattering decreased as $Q$ increased, which evidences that it is of the
magnetic origin.

\begin{figure}[t]
 \includegraphics[scale=0.45]{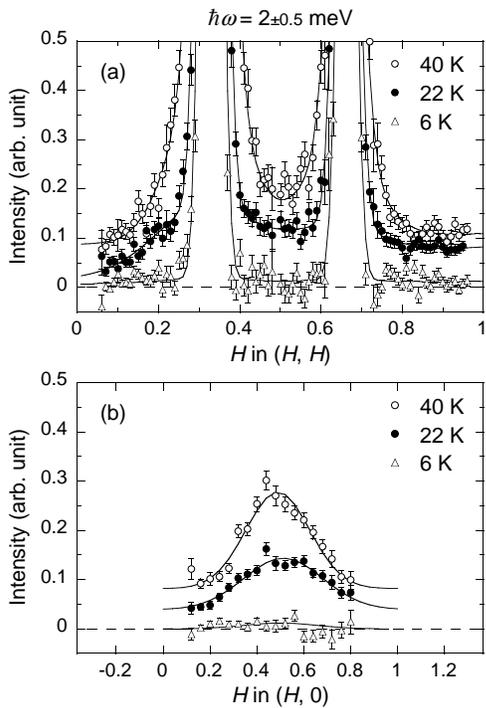}
 \caption{Line profiles of the excitations at 6, 22, and 40~K cut at
 $\hbar\omega = 2 \pm 0.5$~meV and sliced (a) along $(H,H)$ [line A in
 Fig.~\ref{spectra}(j)] with a width of $K = 0\pm 0.02$ in $(-K,K)$ and
 (b) along $(H,0)$ [line B in Fig.~\ref{spectra}(j)] with a width of $K
 = 0\pm 0.02$ in $(-K,2K)$. Solid lines are fits to Gaussians as
 described in the text.}
 \label{SliceTdep}
\end{figure}

To further characterize the two components of the magnetic excitations,
the $T$ dependence of their $Q$ profiles was investigated more
quantitatively. Figure~\ref{SliceTdep} shows the cuts of the spectra in
Figs.~\ref{spectra}(g)--(i) along $(H,H)$ and $(H,0)$. At 22 and 40~K,
the profiles along $(H,H)$ consisted of broad peaks for the diffuse
scattering around sharp peaks for the spin wave excitations at $H=1/3$ and
2/3 [Fig.~\ref{SliceTdep}(a)]. The profiles along $(H,0)$, on the other
hand, exhibited broad single peaks at $H=0.5$
[Fig.~\ref{SliceTdep}(b)]. The intensity of the broad component
decreased as $T$ decreased, and was rarely observed at 6~K. Next, to
parameterize the $T$ dependence, least-square fittings were performed as
follows. For the profiles along $(H,H)$, the intensity ($I$) was fit to
a combination of two sharp Gaussians of the spin wave component, two
broad Gaussians of the diffuse component, and the background ($B$), as
expressed by the following equation:
\[
 I = |f(Q)|^2 \sum_{H_c}
  \left[A_s e^{-\ln{2}\frac{(H-H_c)^2}{\kappa_s^2}}
       + A_d e^{-\ln{2}\frac{(H-H_c)^2}{\kappa_d^2}}
  \right] + B.
\]
Here, $A_s$ ($A_d$) and $\kappa_s$ ($\kappa_d$) are the amplitude and
half width at half maximum (HWHM) of the spin wave component (the
diffuse component), respectively. $H_c$ is the center of each peak ($H_c
\sim 1/3$ and 2/3), and $f(Q)$ is the magnetic form factor $\langle j_0
\rangle$ for Cr$^{3+}$ \cite{IntTable}. The profiles along $(H,0)$ were
fit to single broad Gaussians with the background. Linearly sloping and
constant backgrounds were assumed for the profiles along $(H,H)$ and
$(H,0)$, respectively. The results of the fittings for the 6, 22, and
40~K data are shown as solid lines in Fig.~\ref{SliceTdep}.

\begin{figure}[t]
 \includegraphics[scale=0.45]{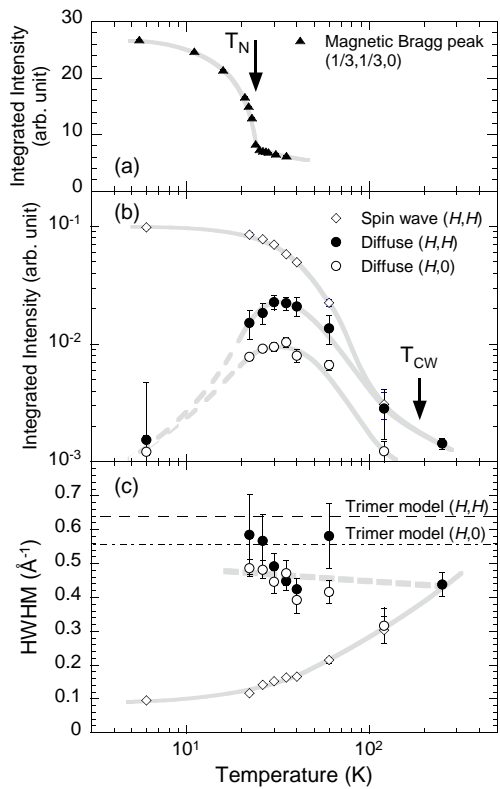}
 \caption{(a) $T$ dependence of the integrated intensity of the magnetic
 Bragg peak $(1/3,1/3,0)$. $T$ dependences of (b) the integrated
 intensities divided by the Bose factor and (c) peak widths (HWHM) of
 the excitation profiles at $\hbar\omega = 2 \pm 0.5$~meV obtained by
 fitting the line profiles as shown in Fig.~\ref{SliceTdep}. In (c),
 HWHM is expressed in \AA$^{-1}$. Open squares, closed circles, and open
 circles show the data for the spin wave component along $(H,H)$, the
 diffuse component along $(H,H)$, and the diffuse component along
 $(H,0)$, respectively. Gray lines are guides to the eye. In (c), the
 peak widths for the trimer model along $(H,H)$ and along $(H,0)$ are
 shown as dashed and dashed-dotted lines, respectively.}
 \label{FitTdep}
\end{figure}

Figures~\ref{FitTdep}(b) and \ref{FitTdep}(c) compile the results of the
$T$ dependence fittings of the integrated intensities and HWHM,
respectively. The integrated intensity was corrected using the Bose
factor $n(\omega,T)+1 = [\exp(\hbar\omega/k_BT)-1]^{-1}+1$. Because of
the weakness of the diffuse components at 6 and 120~K, the values at
30~K were used for the HWHMs along $(H,H)$ at 6 and 120~K and $(H,0)$ at
6~K. At $250~\mathrm{K}$, the intensity of the excitation spectrum was
too weak to distinguish the two components. The profile along $(H,H)$
was then fit to two Gaussians at $H=1/3$ and 2/3.  With decreasing $T$,
the intensity of the spin wave component increased, and was nearly
saturated below $T_N$. The diffuse component further increased with
decreasing $T$, and then decreased after exhibiting a broad maximum at
$T_\mathrm{peak} \sim 35$~K. By studying a powder sample
\cite{kajimoto10}, it was confirmed that a finite amount of the diffuse
component survives above 13~K, but becomes negligible at 6~K. With
respect to the peak widths, the HWHM of the spin wave component
monotonically decreased as $T$ decreased, indicating the development of
the correlation length, although it did not diverge at $T_N$. On the
other hand, that of the diffuse component exhibited no systematic $T$
dependence except for a slight increase.

One of the interesting features of the diffuse magnetic excitations is
that they existed on the lines connecting the K points in the 2D $Q$
space [Figs.~\ref{spectra}(h) and \ref{spectra}(i)]. Diffuse
quasielastic or inelastic scatterings with characteristic $Q$ dependence
have also been observed in several spinels in which the corner-shared
tetrahedra of spins form 3D geometrically frustrated systems
\cite{lee02,kamazawa04,chung05,tomiyasu08,tomiyasu13}. In these cases,
the diffuse scatterings are interpreted as formation of uncorrelated
spin-cluster-like quasiparticles. Thus, the experimental $\mathbf{Q}$
pattern revealed in the present study was compared to the two-body
correlation function of a classical spin cluster model, as in
\cite{lee02,tomiyasu08}. As a model, it was assumed that the neutrons
are scattered by uncorrelated spin trimer units, in each of which three
spins occupy the vertices of a minimum triangle corresponding to the
constituent of the 120$^{\circ}$ long-range antiferromagnetic
order. Figure~\ref{spectra}(k) shows the distribution of the neutron
scattering intensities on the 2D $Q$ space based on this model, in which
the magnetic form factor for Cr$^{3+}$ was considered. The overall
pattern of the observed neutron scattering intensities was well
reproduced by this simple model. The widths of the intensity profiles
along $(H,H)$ and $(H,0)$, as shown in Fig.~\ref{FitTdep}(c) by dashed
and dashed-dotted lines, respectively, were obtained by cutting this
calculated intensity pattern.  Although the observed widths of the
diffuse scattering profiles were smaller than the calculated values,
indicating that the correlation length of the diffuse component was
larger than the size of the trimer, the difference was not very
significant.

Based on the trimer model, the following scenario is proposed for the
development of the spin correlation in {\CCO}. Spins on a triangular
unit of Cr ions begin to develop a 120$^{\circ}$ correlation (trimer)
below $\sim$$T_\mathrm{CW}$ that induces the diffuse scattering of the
spin excitations. With decreasing $T$, a portion of these spins develop
correlation lengths exhibiting 2D spin wave excitations. Simultaneously,
the uncorrelated trimers increase their numbers and coexist with the
matrix exhibiting the spin waves. After saturating around
$T_\mathrm{peak}$, the trimer correlation crosses over to the spin wave
correlation. Although the uncorrelated trimers survive even below $T_N$,
they finally disappear at $T=0$.

The broad maximum of the diffuse component around $T_\mathrm{peak}$
should be related to the broad shoulder structure in $C_\mathrm{mag}$
around $T_N$ \cite{okuda08}. Similar broad shoulder structures in
$C_\mathrm{mag}$ have been widely observed in other $S=3/2$ TLHAs and
kagome-lattice antiferromagnets
\cite{nakatsuji05,olariu06,okuda09,okuda11,takatsu09,ramirez00}, which
have been attributed to the formation of spin singlets \cite{ramirez00}
or clusters \cite{nakatsuji05}. The present result strongly suggests
that the identity of these ``singlets'' or ``clusters'' is the spin
trimers. Furthermore, the crossover behavior in the spin correlations in
{\CCO} is analogous to the unconventional crossover of spin fluctuations
in NaCrO$_{2}$ \cite{olariu06}. The similar decreases in the spin
trimers for {\CCO} should be the origin of the crossover phenomenon in
NaCrO$_{2}$. Thus, we believe that the crossover in the spin
fluctuations accompanied by the uncorrelated trimers is a universal
phenomenon in the development of the spin correlations in 2D TLHA.

Then, what is the origin of the trimer-like correlation in the spin
fluctuations? One of the most fascinating possibilities is to relate the
correlation to the formation of the $Z_2$-vortex
\cite{kawamura84,okubo10,kawamura11}. Because the $Z_2$-vortex predicted
in 2D TLHAs is a vortex of the chirality vectors defined by local
120$^{\circ}$ structures, it results in essentially the same behavior as
the trimer model when the correlation length is short. Recently, a
theoretical study showed that the excitations of the $Z_2$-vortex should
produce diffuse scattering connecting the K points
\cite{okubo10}. Interestingly, the previously predicted $\mathbf{Q}$
pattern of the diffuse scattering looks very similar to those in
Figs.~\ref{spectra}(h) and \ref{spectra}(i). Furthermore, the
$Z_2$-vortex shows a crossover to the spin wave correlations as $T$
decreases \cite{okubo10}. A recent electron spin resonance spectroscopy
study of $A$CrO$_{2}$ further suggested the existence of the
$Z_2$-vortices at $T > T_N$ \cite{hemmida11}. However, it remains
unclear whether the $T$ independent correlation length of the spin
trimers observed in the present study is consistent with the
$Z_2$-vortices. If there is a $T$ region where the correlation length of
the $Z_2$-vortex becomes considerably larger than that of the trimer but
the spin wave correlation is not very developed, it would be possible to
unambiguously distinguish the $Z_2$-vortex. Such a $T$ region may exist
below $T_\mathrm{peak}$, but a more detailed study with finer $T$ steps
and higher statistics is required.

%\section{Conclusion}

In conclusion, we performed an INS study of a single crystal of the 2D
TLHA {\CCO} to investigate the unconventional diffuse component in its
magnetic excitations revealed during a previous study of a powder
sample. We observed that the diffusive scattering shows a characteristic
$Q$ pattern, which is consistent with scattering by spin clusters on
minimum triangular-lattice units. The diffuse component crosses over to
the spin wave component as $T$ decreases, but survives even below $T_N$
without developing its correlation length. The present results suggest
that development of the spin correlation in TLHAs cannot be described
solely by the divergence of the correlation length as in a conventional
magnet, but is accompanied by a decrease in the robust spin clusters.

\begin{acknowledgments}

We thank T. Okubo, H. Kawamura, M. Itoh, Y. Kobayashi, K. Kakurai, and
M. Arai for valuable discussions. The experiments on AMATERAS were
performed with the approval of J-PARC (Proposal Nos.\ 2012A0113 and
2013A0087). This study was supported by JSPS KAKENHI Grant Nos.\ 25400378
and 26800174.

\end{acknowledgments}

\end{document}